\documentclass[11pt]{article}
\usepackage[margin=1in]{geometry}
\usepackage{cprotect}
\usepackage{color}
\usepackage[colorlinks=true,linkcolor=blue,citecolor=blue,urlcolor=blue]{hyperref}
\usepackage{amssymb,amsmath}
\usepackage{amsthm}
\theoremstyle{remark}
\newtheorem{remark}{Remark}

\usepackage{graphicx}
\usepackage{helvet}
\usepackage[font=footnotesize,labelfont=bf]{caption}
\usepackage{longtable}

\usepackage{natbib}
\usepackage{bm}
\newcommand\blfootnote[1]{%
  \begingroup
  \renewcommand\thefootnote{}\footnote{#1}%
  \addtocounter{footnote}{-1}%
  \endgroup
}

\usepackage{setspace}
\usepackage[dvipsnames]{xcolor}
\newcommand{\JJL}[1]{\textcolor{black}{#1}}
\newcommand{\ww}[1]{\textcolor{black}{#1}}
\newcommand{\w}[1]{\textcolor{black}{#1}}
\newcommand{\jl}[1]{\textcolor{black}{#1}}
\newcommand{\add}[1]{\textcolor{black}{#1}}

\DeclareMathOperator*{\argmin}{arg\!min}
\DeclareMathOperator*{\argmax}{arg\!max}

\numberwithin{equation}{section}

\newcommand{\E}{{\rm I}\kern-0.18em{\rm E}}
\newcommand{\Prob}{{\rm I}\kern-0.18em{\rm P}}
\newcommand{\1}{{\rm 1}\kern-0.24em{\rm I}}

\title{Modeling and analysis of RNA-seq data: \\ a review from a statistical perspective}
\author{%
Wei Vivian Li\,$^{1}$ and Jingyi Jessica Li\,$^{1,2,*}$
}
\date{}

\begin{document}

\maketitle

\blfootnote{\\$^{1}$ Department of Statistics, University of California, Los Angeles, CA 90095-1554\\
$^{2}$ Department of Human Genetics, University of California, Los Angeles, CA 90095-7088\\
$^{*}$ To whom correspondence should be addressed.
Email: jli@stat.ucla.edu
}

\begin{abstract}
Background: 
Since the invention of \JJL{next-generation} RNA sequencing (RNA-seq) technologies, they have become a powerful tool to study the presence and quantity of RNA molecules in biological samples \JJL{and \jl{have} revolutionized transcriptomic} studies. The analysis of RNA-seq data at \w{four} different levels (samples, genes, transcripts, and exons) involve multiple statistical and computational questions, some of which remain challenging up to date.

Results: \JJL{We review RNA-seq analysis tools at the \w{sample,} gene, transcript, and exon levels from a statistical perspective.
We also highlight the biological and statistical questions of most practical considerations.}

Conclusion: 
\JJL{The development of statistical and computational methods} for \JJL{analyzing} RNA-seq data has \JJL{made significant advances in the past decade.} However, \JJL{methods developed to answer the same biological question often rely on diverse statical models and exhibit different performance under different scenarios.} This review discusses and compares \JJL{multiple commonly used statistical models regarding their assumptions, in the hope of helping users select appropriate methods as needed, as well as assisting developers for future method development}.
\end{abstract}

\section{Introduction}\label{sec:intro}
RNA sequencing (RNA-seq) uses the next generation sequencing (NGS) technologies to reveal the presence and quantity of RNA molecules in biological samples. Since its invention, RNA-seq has revolutionized transcriptome analysis in biological research. RNA-seq does not require any prior knowledge on RNA sequences, and its high-throughput manner allows for genome-wide profiling of transcriptome landscapes \cite{wang2009rna, zhao2014comparison}. Researchers have been using RNA-seq to \JJL{catalog all transcript species, such as messenger RNAs (mRNAs) and long non-coding RNAs (lncRNAs)}, to determine the transcriptional structure of genes, and to quantify the dynamic expression patterns of \JJL{every} transcript under different biological conditions \citep{wang2009rna}. 

\begin{figure}[htb!]
\begin{center}
\includegraphics[width = \textwidth]{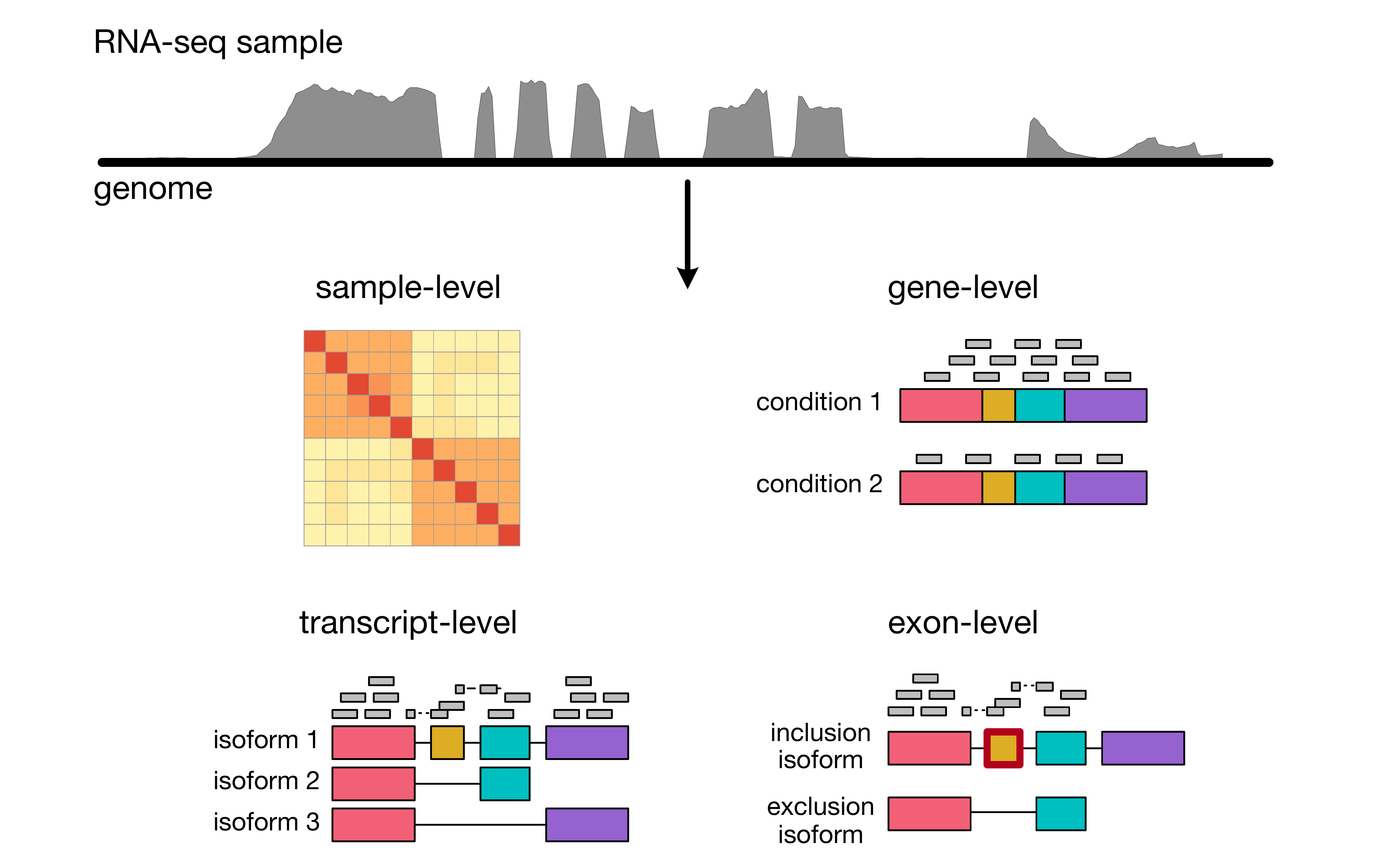}
\end{center}
\caption{\label{fig: outline}
\w{RNA-seq analyses at four different levels: sample-level, gene-level, transcript-level, and exon-level. In the sample-level analysis, the results are usually summarized into a similarity matrix, as introduced in Section \ref{sec:trom}. Taking a 4-exon gene as an example, the gene-level analysis summarizes the \jl{counts of RNA-seq reads mapped to genes} in samples of different conditions, and it subsequently compares \jl{genes'} expression levels calculated based on read counts; the transcript-level analysis focuses on reads \jl{mapped} to different isoforms; the exon-level analysis mostly considers the reads \jl{mapped} to or skipping the exon of interest (the \jl{yellow exon marked by a red box} in this example).
}}
\end{figure}

Due to the popularity of RNA-seq technologies and the increasing needs to analyze large-scale RNA-seq datasets, more than \ww{two thousand} computational tools have been developed in the past ten years to assist the visualization, processing, analysis, and interpretation of RNA-seq data. The two most computationally intensive steps are data processing and analysis. In data processing, for organisms with reference genomes available, short RNA-seq reads (fragments) are aligned (or mapped) to the reference genome and converted into genomic positions; for organisms without reference genomes, \textit{de novo} transcriptome assembly is needed.
Regarding the reference-based alignment, the RNA-seq Genome Annotation Assessment Project (RGASP) Consortium has conducted a systematic evaluation of mainstream spliced alignment programs for RNA-seq data \citep{engstrom2013systematic}. We refer interested readers to this paper and do not discuss these alignment algorithms here, as statistical models are not heavily involved in the alignment step. 
In this paper, we focus on the statistical questions engaged in RNA-seq data analyses, assuming reads are already aligned to the reference genome. 
Depending on the biological questions to be answered from RNA-seq data, we categorize RNA-seq analyses at \w{four} different levels, which require three different ways of RNA-seq data summary.
\jl{Sample-level analyses (e.g., sample clustering) and gene-level analyses (e.g., identifying differentially expressed genes \citep{soneson2013comparison} and constructing gene co-expression networks \citep{giorgi2013comparative})} mostly require gene read counts, i.e., how many RNA-seq reads are mapped to each gene.
Note that we refer to a transcribed genomic region as a ``gene" throughout this review, and a multi-gene family (multiple transcribed regions that encode proteins with similar sequences) are referred to as multiple ``genes".
Transcript-level analyses, such as RNA transcript assembly and quantification \citep{kanitz2015comparative}, often need read counts of genomic regions within a gene, i.e., how many RNA-seq reads are mapped to each region and each region-region junction, or even the exact position of each read.
Exon-level analyses, such as identifying differential exon usage \citep{tourasse2017quantitative}, usually require read counts of exons and exon-exon junctions.
As these \w{four levels of analysis \jl{use}} different statistical and computational methods, we will review the key statistical models and methods widely used at each level of RNA-seq analysis (Figure \ref{fig: outline}), with \JJL{an} emphasis on the identification of differential expression and alternative splicing patterns, two of the most common goals of RNA-seq experiments.

This review does not aim to exhaustively enumerate all the existing computational tools designed for RNA-seq data, but to discuss the strategies of statistical modeling and application scopes of typical \JJL{methods} for RNA-seq analysis. 
We refer readers to \cite{wang2009rna,jessica15} for an introduction to the development of RNA-seq technologies\w{, and \citep{seqc2014comprehensive} for} a comprehensive assessment of RNA-seq with a comparison to microarray technologies and other sequence-based platforms by the Sequencing Quality Control (SEQC) project. For considerations in experimental designs and more recent advances in analyzing tools, we refer readers to \cite{conesa2016survey}.

\section{\w{Sample-level analysis: transcriptome similarity}}\label{sec:trom}

The availability of numerous public RNA-seq datasets has created an unprecedented opportunity for researchers to compare multi-species transcriptomes under various biological conditions.
Comparing transcriptomes of the same or different species can reveal molecular mechanisms behind important biological processes, and help one understand the conservation and differentiation of these molecular mechanisms in evolution. 
\w{ 
\jl{Researchers need similarity measures to directly evaluate the similarities of different samples (i.e., transcriptomes) based on their genome-wide gene expression data summarized from RNA-seq experiments. Such similarity measures are useful for outlier sample detection, sample classification, and sample clustering analysis.} When \jl{samples represent individual cells, similarity measures may be used to identify} rare or novel cell types.
\jl{In addition to} gene expression, it is also possible to \jl{evaluate transcriptome} similarity based on alternative splicing events \citep{gao2017correspondence}.}
Correlation analysis is a classical approach to measure  \jl{transcriptome similarity} of biological samples \citep{arbeitman2002gene,necsulea2014evolution}. The most commonly used measures are Pearson and Spearman correlation coefficients. The analysis starts by calculating pairwise correlation coefficients of normalized gene expression between any \jl{two} biological samples, resulting in a correlation matrix. Users can visualize the correlation matrix (usually as a heatmap) to interpret \jl{the pairwise transcriptome similarity of biological samples}, or they may use the correlation matrix in downstream analysis such as \jl{sample} clustering.

However, a caveat of using correlation analysis to infer transcriptome similarity is that the existence of housekeeping genes would inflate correlation coefficients. Moreover, correlation measures rely heavily on the accuracy of gene expression measurements and are not robust when the signal-to-noise ratios are relatively low. \jl{Therefore, we have developed an alternative} transcriptome overlap measure \texttt{TROM} \citep{li2017trom} to find sparse correspondence of transcriptomes in the same or different species. The \texttt{TROM} method compares biological samples based on their ``associated genes'' instead of the whole gene population, thus leading to a more robust and sparse transcriptome similarity result than that of the correlation analysis. TROM defines the associated genes of a sample as the genes that have $z$-scores (normalized expression levels across samples per gene) greater or equal than a systematically selected threshold. Pairwise \texttt{TROM} scores are then calculated by an overlap test to measure the similarity of associated genes \jl{for} every pair of samples. The resulting  \texttt{TROM} score matrix has the same dimensions as the correlation matrix, with rows and columns corresponding \jl{to} the samples used in the comparison, and \jl{the \texttt{TROM} score matrix} can be easily visualized or incorporated into downstream analyses.

\w{Aside from the correlation coefficients and the \texttt{TROM} scores, there are other statistical measures useful for measuring transcriptome similarity in various scenarios. First, partial correlation \jl{can be used to measure sample similarity after eliminating the part of the sample correlation} attributable to another variable such as batch \jl{effects} or experimental \jl{conditions} \citep{de2004discovery}. Second, with evidence of \jl{a} non-linear association between RNA-seq samples, it is suggested to use \jl{measures that can  capture} non-linear dependences, such as the mutual information (MI). Similarly, one may consider using the conditional mutual information (CMI) \citep{wyner1978definition} or \jl{partial mutual information (PMI)} \citep{zhao2016part} to remove the \jl{effects} of other confounding variables.
In addition to \jl{the direct calculation of the sample similarity matrix by applying a similarity measure to the high-dimensional gene expression data}, sometimes it is helpful to visualize the gene expression data and investigate \jl{the sample similarities after} dimension reduction. \jl{Popular dimension reduction} methods include principal component analysis (PCA), \jl{t-stochastic neighbor embedding (t-SNE)} \citep{maaten2008visualizing}, and multidimensional scaling (MDS) \citep{kruskal1978multidimensional}.
}

\section{Gene-level analysis: gene expression dynamics}\label{sec:genes}


RNA-seq technologies enable the measurement and comparison of genome-wide gene expression patterns across different samples without the restriction of known genes, which are required by microarray experiments. The profiling of gene expression patterns is the key to \JJL{investigating new} biological processes in \jl{various tissues and cells of different} organisms. \JJL{A common but important question} in a large cohort of biological studies is \JJL{how} to compare gene expression levels across different experimental conditions, \JJL{time points,} tissue and cell types, or even species. 
When \JJL{a biological study concerns} two different biological conditions, \JJL{differential gene expression (DGE) analysis is useful for comparing RNA-seq samples of the two conditions}. When the number of biological conditions far exceeds two, \JJL{though DGE analysis can still be used to compare samples in a pairwise manner, a more useful way is to simultaneously measure the transcriptome similarity of multiple samples\jl{, as we have described in Section 2: sample-level analysis}.}

\subsection{Differential gene expression analysis}

The main \JJL{approach to comparing two biological conditions} is to find \JJL{``differentially expressed'' (DE)} genes. A gene is defined as DE if it is transcribed into different amounts of mRNA molecules per cell under the two conditions \citep{evans2017selecting}. However, since we do not observe the true amounts of mRNA molecules, statistical tests are \jl{principled approaches that} help biologists \JJL{understand} to what extent a gene is DE.

It is commonly acknowledged that normalization is a crucial step \jl{prior to} \JJL{DGE} analysis due to \JJL{the existence of batch effects, which could arise from different sequencing depths or} various protocol-specific biases in different experiments \citep{bullard2010evaluation}. The reads per kilobase per million mapped reads (RPKM) \citep{mortazavi2008mapping}\jl{, the fragments per kilobase per million mapped reads (FPKM) \citep{trapnell2009tophat}, and the transcripts per million mapped reads (TPM) \citep{li2011rsem} are the three} most frequently used units for gene expression measurements from RNA-seq data, and they remove the effects of total sequencing depths and gene lengths. \jl{The main difference between RPKM and FPKM is that the former is a unit based on single-end reads, while the latter is based on paired-end reads and counts the two reads from the same RNA fragment as one instead of two. The difference between RPKM/FPKM and TPM is that the former calculates sample-scaling factors before dividing read counts by gene lengths, while the latter divides read counts by gene lengths first and calculates sample-scaling factors based on the length-normalized read counts. If researchers would like to interpret gene expression levels as the proportions of RNA molecules from different genes in a sample, TPM has been suggested as a better unit than RPKM/FPKM \citep{wagner2012measurement}. Even though in these units, gene expression data may still contain protocol-specific biases \citep{dillies2013comprehensive}, and further normalization is often needed.} There are \jl{two main} categories of normalization methods: distribution-based and gene-based. Distribution-based normalization methods aim to make the distribution of all or most gene expression \JJL{levels} similar across different \JJL{samples}, and such methods include \JJL{the quantile normalization} \citep{bolstad2003comparison}, \texttt{DESeq} \citep{anders2010differential}, and \texttt{TMM} \cite{robinson2010scaling}. Gene-based normalization methods aim to make non-DE genes or housekeeping genes have the same expression levels \JJL{in different samples}, and such methods include \JJL{a method by \textit{Bullard et al.} \citep{bullard2010evaluation} and \texttt{PoissonSeq} \citep{li2012normalization}.}
For a comprehensive comparison \JJL{of the assumptions and performance of these normalization} methods, we refer readers to \cite{evans2017selecting, bullard2010evaluation, rapaport2013comprehensive}.

\begin{figure}[htb!]
\begin{center}
\includegraphics[width = \textwidth]{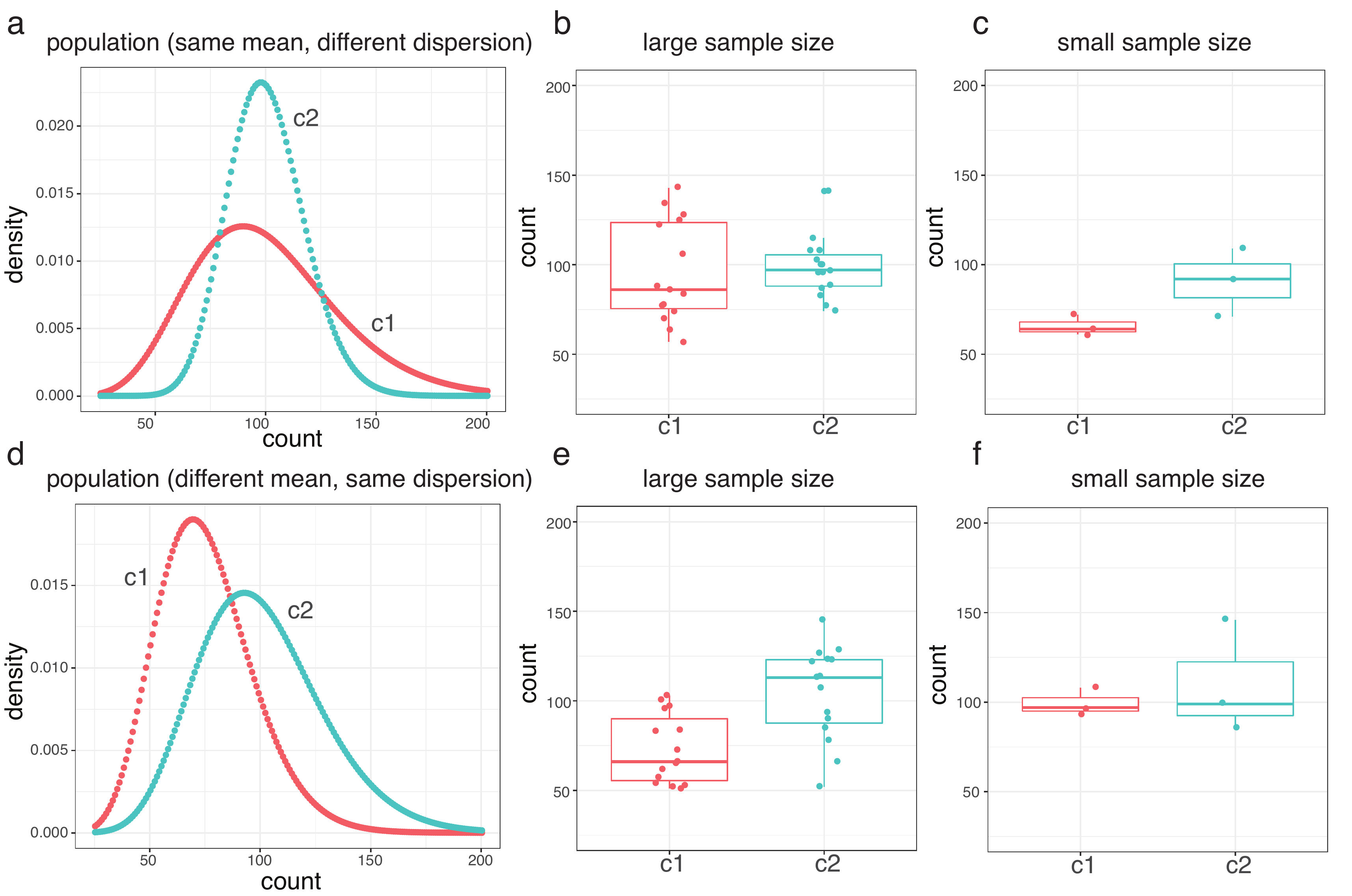}
\end{center}
\caption{\label{fig: de_nb}
\w{Illustration of read counts as samples drawn from \jl{unobservable populations}.
\textbf{a}: The population read count distribution of \jl{a} hypothetical gene 1 in \jl{conditions} c1 and c2, based on the NB model. The two population distributions have the same mean \jl{parameter} but different dispersion \jl{parameters.}
\textbf{b-c}: The observed read counts of gene 1 are independent samples \jl{(\textbf{b}: with large sample sizes; \textbf{c}: with small sample sizes)} drawn from the \jl{two unobservable distributions}. When \jl{the} sample size is small, \jl{a statistical test about whether the two samples have the same population mean will possibly lead to a false positive} result (gene 1 \jl{found as} a DE gene).
\textbf{d}: The population read count distribution of \jl{a} hypothetical gene 2 in 
\jl{conditions} c1 and c2, based on the NB model. The two population \jl{distributions} have different \jl{mean parameters} but the same dispersion \jl{parameter}.
\jl{\textbf{e-f}: The observed read counts of gene 2 are independent samples \jl{(\textbf{e}: with large sample sizes; \textbf{f}: with small sample sizes)} drawn from the two unobservable distributions. When the sample size is small, a statistical test about whether the two samples have the same population mean will possibly lead to a false negative result (gene 2 found as a non-DE gene).}
}}
\end{figure}

\JJL{How to form a proper statistical hypothesis test is the core question in the development of a DGE method.} Most existing methods use the Poisson distribution \citep{bloom2009measuring} or the Negative Binomial (NB) distribution \citep{anders2010differential,robinson2010edger,hardcastle2010bayseq} to model \jl{the} read counts of an individual gene \jl{in different samples} \w{(Figure \ref{fig: de_nb}\jl{a, d})}. \jl{In our discussion here, we focus on the NB distribution because it is commonly used to account for the observed} over-dispersion of RNA-seq read counts. Throughout this section, we consider two biological conditions $k = 1, 2$, each with $J_k$ samples. $Y_{k,ij}$ denotes the read count of gene $i$ in the $j$th sample of condition $k$. The basic assumption is that
\begin{align}
\begin{split}\label{eq:negb}
Y_{k,ij}&\sim \text{NB}(\text{mean} = s_{kj}\theta_{ki},\ \text{dispersion}=\phi_i),
\end{split}
\end{align}
where $s_{kj}$ is the size factor of the $j$th sample \jl{of} condition $k$, $\theta_{ki}$ is the \jl{true expression level} of gene $i$ \jl{under} condition $k$, and $\phi_i$ is the dispersion of gene $i$. 
\w{It is necessary to consider the size factor $s_{kj}$ \jl{because it accounts for the fact that different samples usually have different numbers of sequenced reads}. The dispersion \jl{parameter $\phi_i$ controls the variability of the expression levels of gene $i$ across biological samples. The estimation of the parameters $s_{kj}$, $\theta_{ki}$ and $\phi_i$ is the key step to investigating the differential expression of gene $i$ between the two conditions}.}
Bayesian modeling is often used, and prior distributions and relationships of $s_{kj}$, $\theta_{ki}$, and $\phi_i$ are often assumed. Note that assuming $s_{kj}$ being independent of gene $i$ simplifies the problem, but it can be advantageous to calculate gene-specific factors $s_{k,ij}$ to account for technical biases dependent on \JJL{gene-specific GC contents or gene lengths} \citep{love2014moderated}.
\JJL{The DGE analysis is carried out by testing} 
\begin{align}
\begin{split}
H_0: \theta_{1i} = \theta_{2i}\ \text{vs.}\ H_1: \theta_{1i} \neq \theta_{2i}
\end{split}
\end{align}
for each gene $i$. 

Starting from \JJL{the model} (\ref{eq:negb}), most methods include six steps. First, they estimate \jl{$\theta_{ki}$ and $\phi_i$} for each gene. 
\jl{Under the NB distribution, the dispersion parameter characterizes the mean-variance relationship, consistent with the observation that genes with similar true expression levels exhibit similar variances \citep{robinson2010edger,love2014moderated}.
When the sample sizes are small (Figure \ref{fig: de_nb}c, f), one may consider using  shrinkage estimation of $\phi_i$'s to borrow information across genes or to incorporate prior knowledge, for the purpose of obtaining more robust results \citep{yu2013shrinkage}.}
Second, they construct a test statistic based on the estimators to reflect the mean difference between the two conditions. Third, they derive the null distribution of the test statistic under $H_0$. Fourth, they calculate the \ww{observed value of test statistic} for each gene. Fifth, they convert the \ww{observed values of test statistics} into $p$-values based on the null distribution. Sixth, they perform multiple test correction on the $p$-values to determine a reasonable threshold, and the genes with $p$-values under that threshold would be called as DE. 

For example, \texttt{edgeR} \citep{robinson2010edger} first estimates the \JJL{dispersion parameters} using a conditional maximum likelihood, and it then \JJL{develops} a test analogous to the Fisher's exact test. \texttt{DESeq2} \citep{love2014moderated} adds a layer to the model by estimating \JJL{$(\theta_{2i}-\theta_{1i})$} using \JJL{a} generalized linear model with a logarithmic link function, \JJL{$Y_{k,ij}$ as the response variable, and the condition as a binary predictor (i.e., whether the condition $k = 2$). This generalized linear model setup can easily incorporate the information on experimental design as additional predictors.} In the testing step, \texttt{DESeq2} transforms the problem \JJL{into testing if the condition predictor has significant effects on the logarithmic fold change of gene expression, which is equivalent to testing whether $\theta_{2i}-\theta_{1i}=0$}. 
\texttt{EBSeq} \citep{leng2013ebseq} and \texttt{ShrinkSeq} \citep{van2013bayesian} are also based on \JJL{the model} (\ref{eq:negb}), but under \JJL{a} Bayesian framework they \JJL{use hyper-parameters to borrow information across genes, and they} directly calculate the posterior probability of a gene being differentially expressed \JJL{, i.e., $\Prob\left(\theta_{1i}\neq \theta_{2i} \mid Y_{1,i1}, \ldots, Y_{1,iJ_1}, Y_{2,i1}, \ldots, Y_{2,iJ_2}\right)$.} 

\add{There are other DGE methods that do not assume the negative Binomial distribution as in the model (\ref{eq:negb}) but take a different approach by assuming that $\log(Y_{k,ij})$ follows a Normal distribution, which has much more tractable mathematical theory than count distributions (such as the negative Binomial distribution) have. For example, the \texttt{voom} method \cite{law2014voom} estimates the mean-variance relationship of $\log(Y_{k,ij})$ and generates a precision weight for each observation. Then \texttt{voom} inputs $\log(Y_{k,ij})$ and precision weights into the limma empirical Bayes analysis pipeline \cite{smyth2005limma}, which is designed for microarray data and has multiple modeling advantages: using linear modeling to analyze complex experiments with multiple treatment factors, using quantitative weights to account for variations in the precision of different observations, and using empirical Bayes methods to borrow information across genes. Another method \texttt{sleuth} \citep{pimentel2017differential} is applicable to finding both  differentially expressed genes and transcripts between two conditions. Here we describe \texttt{sleuth} in the context of DGE analysis. \texttt{Sleuth} uses a linear model with $\log(Y_{k,ij})$ as the response variable, and \texttt{sleuth} decomposes the variance of $\log(Y_{k,ij})$ into three components: the variance explained by the condition predictor (whose coefficient is the parameter of interest and indicates differential expression if non-zero), the variance of ``biological noise" (which accounts for the variance of true gene expression across samples of the same condition), and the variance of ``inferential noise" (which accounts for the additional variance of observed gene expression due to the uncertainty in gene expression estimation). \texttt{Sleuth} assumes both the ``biological noise" and the ``inferential noise" follow independent zero-mean Gaussian distributions. For every gene, \texttt{sleuth} estimates the variance of the ``inferential noise"  by bootstrapping RNA-seq reads to estimate the variance of the expression estimates of that gene. Accurate estimation of the variance of the ``inferential noise" allows better estimation of the null distribution of the test statistic, i.e., the estimator of the coefficient of the condition predictor, in the third step, thus leading to more accurate estimates of the $p-$values and false discovery rates.}

\begin{remark}
A common scenario is that a study only includes a small number of RNA-seq replicates \citep{schurch2016many}. Even though most methods introduced in this section \JJL{are technically applicable} to data with as few as two replicates per condition, there is no guarantee of good performance for these methods with a small number of replicates. \ww{In fact}, it was observed that many methods did not have a good control on false discovery rates (FDR) under this scenario \citep{schurch2016many} \w{(Figure \ref{fig: de_nb})}.
We suggest users carefully check or consult a statistician if the assumptions of a method are reasonable for their study before using the method, as a way to reduce the chance of misusing statistics.
\end{remark}

\begin{remark}\label{remark:de_compare}
Comparisons of \JJL{DGE} methods show that none of the methods is optimal in all circumstances, and methods can produce very different results (regarding both \JJL{the ranking and number of DE genes}) on the same dataset  \citep{soneson2013comparison,rapaport2013comprehensive}. In \JJL{some} applications, \JJL{users are more concerned about the ranking of DE genes than the resulting $p$-values of genes, especially when setting a reasonable threshold on the $p$-values is difficult}. 
\JJL{In other applications where thresholding on $p$-values is required to control the probability that a gene is falsely discovered as DE}, users need to address the multiple-testing issue\JJL{, as testing for tens of thousands of genes simultaneously could lead to a large number of false discoveries even at a small $p$-value threshold}. 
\JJL{Common approaches to address the multiple-testing issue} include the Bonferroni correction \citep{neyman1928use}, the Holm-Bonferroni method \citep{holm1979simple}, and the \JJL{Benjamini-Hochberg false discovery rate (FDR)} correction \citep{benjamini1995controlling}, with \JJL{a decreasing level of \w{conservatism}}. The first two methods aim to control the family-wise error rate (the probability of making one or more false discoveries), while the third method aims to control the expected proportion of false discoveries \JJL{among the discoveries}.
\end{remark}

\begin{remark}
In studies where researchers are interested in temporal dynamics of transcriptomes, RNA-seq data are produced at multiple time points of the same tissue or cell type. To identify the genes whose expression levels along the time course change significantly between two conditions, a previous approach \texttt{maSigPro} \citep{nueda2017identification}
is based on a linear model where gene expression level is modeled as the response variable, while time points and conditions are considered as predictors. The identification of DE genes is then formulated as the problem of testing whether the condition variable has non-zero coefficient for each gene. Another previous work based on microarray data provided a two-sample multivariate empirical Bayes statistic (MB statistic) for replicated microarray time course data \cite{tai2006multivariate}. The MB statistic can be used to test the null hypothesis that the expected temporal expression profiles of one gene under two conditions are the same, and it is thus a criterion to rank genes in the order of evidence of nonzero mean difference between two conditions, incorporating the correlation structure \jl{of} time points, moderation, and replication.  
\end{remark}

\subsection{\w{Gene co-expression network analysis}}

\w{A gene co-expression network (GCN) is an undirected graph, where nodes correspond to different genes, and edges connecting the nodes denote the co-expression relationships between genes. GCNs can help people learn the functional relationships between genes and infer and annotate the functions of unknown genes. To the best of our knowledge, the first GCN analysis on a genome-wide scale across multiple organisms was completed in 2003, enabled by the availability of \jl{high-throughput} microarray data \citep{stuart2003gene}. One of the most commonly used GCN analysis \jl{methods, \texttt{WGCNA}, was initially developed for microarray data, but \texttt{WGCNA} can be applied to} normalized RNA-seq data \citep{langfelder2008wgcna}.
\w{It is popularly applied to gene expression datasets to detect gene clusters and modules and investigate gene connectivity by analyzing correlation networks.}
Here we introduce the GCN \jl{methods} based on the framework proposed in \citep{zhang2005general}. We denote the gene expression matrix as $\bm X_{N\times J}$, where the $N$ rows represent genes\jl{,} and \jl{the} $J$ columns represent \jl{samples}. The $N$ genes are considered as $N$ nodes in the co-expression network. The first step is to construct a symmetric adjacency matrix \jl{$\bm A_{N\times N}$}, where $A_{ij}$ is \jl{a} similarity score \jl{in the range from $0$ to $1$} between \jl{genes} $i$ and $j$. $A_{ij}$ measures the level of concordance between gene expression vectors $\bm X_{i\cdot}$ and $\bm X_{j\cdot}$. As discussed in Section \ref{sec:trom}, the similarity measure can be calculated based on the correlation coefficients or the mutual information measures\jl{, depending on the type of gene co-expression relationships interested} in the analysis. The elements in the adjacency matrix only consider each pair of genes when evaluating their similarity in expression profiles. However, it is important to consider the relative connectedness of gene pairs with respect to the entire network in order to detect co-expression gene modules. Therefore, one needs to calculate the topological overlap matrix \jl{$\bm T_{N\times N}$}, where $T_{ij}$ is the topological overlap between node $i$ and $j$. One such example used in previous studies is \citep{ravasz2002hierarchical}:
$$T_{ij} = \frac{\sum_{k=1}^{N}A_{ik}A_{kj}+A_{ij}}{\min\{\sum_{k=1}^NA_{ik},\sum_{k=1}^NA_{jk}+1-A_{ij}\}}.$$
The final distance between nodes $i$ and $j$ is defined as $d_{ij} = 1- T_{ij}$. Clustering methods can then be applied to search for gene modules based on the \jl{resulting} distance matrix.
The identified gene modules are of great biological interests in many applications. For example, the modules can serve as a prioritizer to evaluate functional relationships between known disease genes and candidate genes \citep{oti2008conserved}. Gene modules can also be used to detect regulatory genes and study the regulatory mechanisms in various organisms \citep{segal2003module}.
}

\section{Transcript-level analysis: transcript reconstruction and quantification}\label{sec:txs}
\begin{figure}[htb!]
\begin{center}
\includegraphics[width = \textwidth]{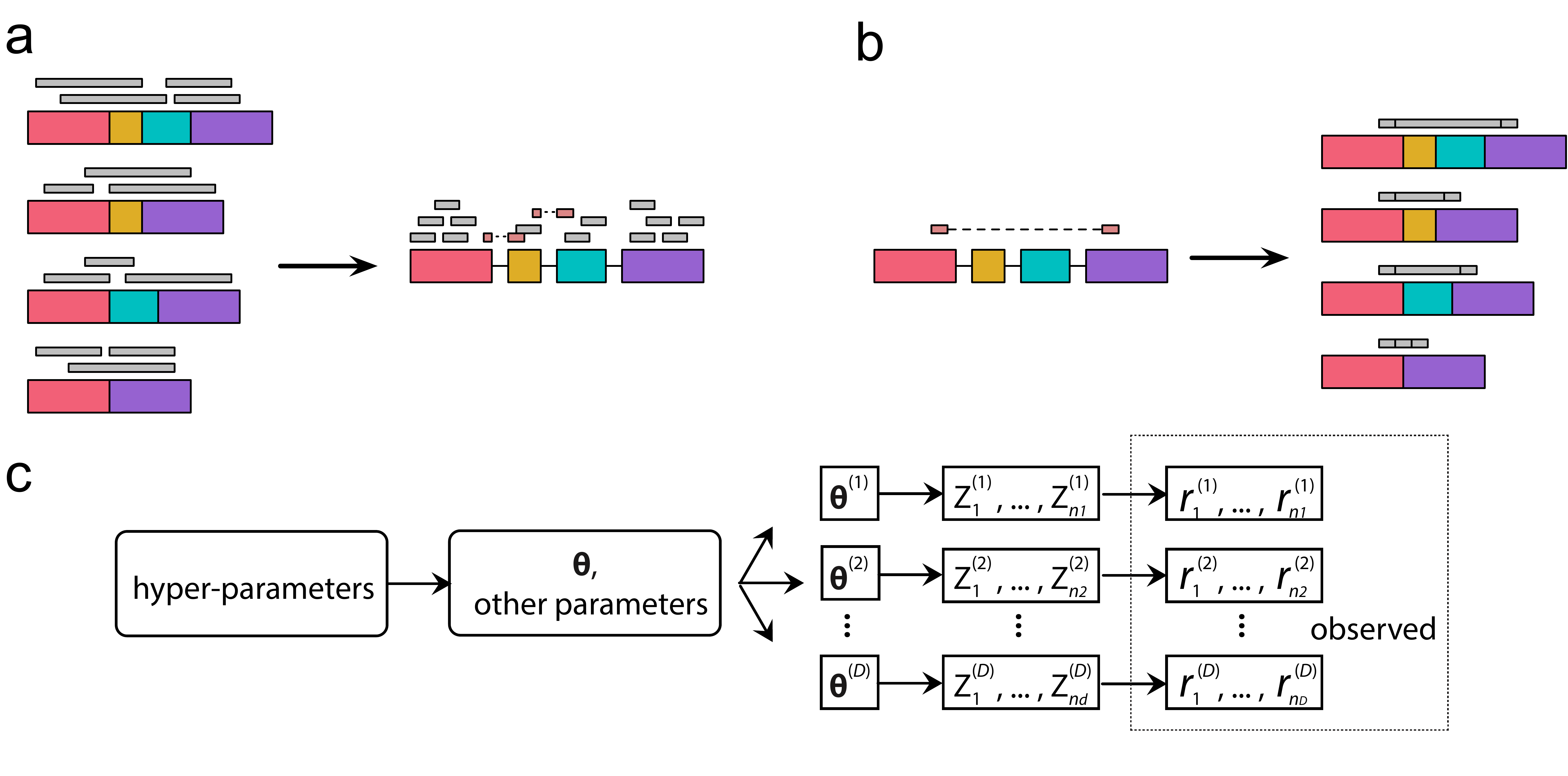}
\end{center}
\caption{\label{fig: txlevel}
\w{\textbf{a}: Taken this 4-exon gene as an example, the observed RNA-seq reads \jl{were sequenced from fragments of the true but unobservable isoform, which contains all the four exons}. The read length is fixed in each experiment, but the fragment lengths can vary. 
Since only the two ends of each fragment are sequenced \jl{as paired-end reads}, this leads to information loss in RNA-seq experiments.
\textbf{b}: Given the paired-end reads mapped to the 4-exon gene (one end mapped to the first exon and the other end mapped to the fourth exon), the inferred fragment length could be different when assuming different isoform origin of the read.
\textbf{c}: An example Bayesian framework to estimate the population isoform \jl{proportions} $\bm \theta$ of a gene given $D$ samples. $\bm \theta^{(1)},\dots,\bm \theta^{(D)}$ are considered as the realization of $\bm \theta$ in $D$ samples. $Z_i^{(d)}$ denotes the isoform origin of read \jl{$r_i^{(d)}$} in sample $d$. Only the reads \jl{$r_i^{(d)}$}'s are observed information\jl{, other random variables are hidden, and parameters need estimation.}
}}
\end{figure}

An \JJL{important use} of RNA-seq data is to recover full-length \JJL{mRNA} transcript structures and expression levels based on short RNA-seq reads. This application involves two major tasks\JJL{.} 
The first task, to identify novel transcripts in RNA-seq samples, is commonly referred to as transcript/isoform reconstruction, discovery, assembly, or identification. This is one of the most challenging problems in this area due to the large searching space of candidate isoforms (especially for complex genes) and limited information contained in short reads \w{(Figure \ref{fig: txlevel}a)}.
The second task, to estimate the expression of known or newly discovered transcripts, is usually referred to as transcript/isoform quantification or abundance estimation.
In recent years, it is a common practice to combine the two tasks into one step\JJL{,} and many popular computational tools simultaneously perform transcript reconstruction and quantification \citep{canzar2016cidane}. This is usually achieved by estimating the expression levels of all the candidate isoforms with penalty or regularity constraints, and the \w{resulting} isoforms with non-zero \JJL{estimated} expression are treated as reconstructed isoforms.
Therefore, we introduce these two tasks together in this review, as they can be tackled by the same statistical framework in many existing tools. We focus on the basic models that are commonly used by multiple methods\JJL{, while selectively introducing} characteristics of individual methods. These models are generally annotation-based and assume that \JJL{a} reference genome is available for the organism of interest.

The transcript reconstruction and quantification \JJL{are performed separately for individual genes}, so the following discussion applies \ww{to} one gene. Throughout this section, we \JJL{index the isoforms of a gene} as $\{1,2,\dots,J\}$. In the reconstruction setting, $J$ is the total number of candidate isoforms to be considered; in the quantification setting, $J$ is \JJL{the number of annotated (or newly discovered) isoforms to be quantified}.
We \JJL{index the exons of the gene} as $\{1,2,\dots,I\}$.
Suppose \JJL{that a total of $n$ (single-end or paired-end) reads} are mapped to the gene\JJL{,} and they are denoted as $\bm R = \{r_1, r_2,\dots,r_n\}$.
The goal of most methods is to estimate $\JJL{\bm \Theta}=(\theta_1,\theta_2,\dots,\theta_J)^T$, where
\begin{align}
\begin{split}
\theta_j &= \text{fraction of isoform}\ j\\
&= \Prob\text{(\JJL{a random} read is from isoform}\ j).
\end{split}
\end{align}

\subsection{Likelihood-based \JJL{methods}}

The first type of \JJL{transcript quantification methods estimates transcript abundance by maximizing the likelihood or the posterior based on a statistical model}. These \JJL{methods} are flexible and can be easily modified to incorporate \JJL{prior biological information into the posterior to improve quantification accuracy}. The \JJL{statistical models are} further divided \ww{into} \add{three categories: region-based, read-based, and fragment-based models}.

Region-based models summarize the read counts based on the genomic regions of interest, such as exons and exon-exon junctions. Suppose \JJL{that} $S$ is the index set that denotes all \JJL{the} regions of interest\JJL{. Read} counts can be summarized as $\bm X = \{X_s \mid s \in S\}$, where $X_s$ \JJL{is} the total number of reads mapped to region $s$. The basic model assumes that $X_s$ follows a Poisson distribution with parameter $\lambda_s$. \JJL{Given} the structures of isoforms and their compatibility with the regions, it is \JJL{reasonable to assume} $\lambda_s$ as a linear function of \JJL{the $\theta_j$'s}: $\lambda_s = \sum_{j=1}^J a_{sj}\theta_j$. The likelihood function can then be derived\JJL{, and the task of estimating $\bm\Theta$} reduces to a maximum likelihood estimation (MLE) problem:
\begin{align}\label{eq:4.2}
\begin{split}
L(\bm \Theta|\bm X) & = \prod_{s \in S}\frac{e^{\JJL{-\lambda_s}}\lambda_s^{X_s}}{X_s!} \JJL{= \prod_{s \in S}\frac{\exp\left\{-\sum_{j=1}^J a_{sj}\theta_j\right\}\left(\sum_{j=1}^J a_{sj}\theta_j\right)^{X_s}}{X_s!}}\,,\\
\hat{\bm \Theta}&= \JJL{\left(\hat\theta_1,\hat\theta_2,\dots,\hat\theta_J\right)^T} = \argmax_{\bm \Theta}\sum_{s \in S}\log L(\bm \Theta | \bm X)\,.
\end{split}
\end{align}
\add{The first isoform quantification method \citep{jiang2009statistical} uses a region-based model.}

\add{In contrast to region-based models}, read-based methods directly \JJL{use the likelihood as a product of the probability densities of individual reads instead of first summarizing reads into region counts.}
\begin{align}\label{eq:4.3}
\begin{split}
L(\bm \Theta | \JJL{\bm R}) & = \prod_{i = 1}^n \JJL{p}(r_i|\JJL{\bm\Theta})\\
&= \prod_{i = 1}^n \sum_{j=1}^J \JJL{p}(r_i| \text{isoform}\ j)\;\theta_j\\
&= \prod_{i = 1}^n \sum_{j=1}^J \JJL{p}(s_i| \text{isoform}\ j)\;\JJL{p}(\ell_{ij} | \text{isoform}\ j)\;\theta_j\,,
\end{split}
\end{align}
where $s_i$ is the starting position and $\ell_{ij}$ is the read length (\JJL{for single-end reads}) or fragment length (\JJL{for paired-end reads}) of read $r_i$ \w{if it belongs to isoform $j$ (Figure \ref{fig: txlevel}b)}. While many methods do not explicitly state it, they assume that the $s_i$ and $\ell_{ij}$ are independent in the above model. 
\w{If the two ends of read $i$ are mapped to the same exon or two neighboring exons, \jl{its corresponding fragment length can be determined and remains the same for all its} compatible isoforms.
\jl{Otherwise, the corresponding fragment length $l_{ij}$ of read $i$ could be different for different compatible isoform $j$} (Figure \ref{fig: txlevel}b). 
Even though each read has the same weight in the likelihood model, the \jl{reads that are mapped to two non-neighboring exons} play a critical role in the detection of splicing junctions and \jl{the reconstruction of full-length transcripts}.}
\texttt{Cufflinks} \citep{trapnell2010transcript}, \texttt{eXpress} \citep{roberts2013streaming}, \texttt{RSEM} \citep{li2011rsem}\add{, and \texttt{Kallisto} \citep{bray2016near}} all adapted or extended the above model in their quantification step, and they mainly differ in how they model $p(s_i| \text{isoform}\ j)$ and $p(l_i| \text{isoform}\ j)$ to incorporate \jl{sequencing} bias adjustment. \add{One feature distinguishing \texttt{Kallisto} from the other methods is that \texttt{Kallisto} speeds up the processing by pseudoaligning the reads and circumventing the computation costs of exact alignment of individual bases. To estimate $\Theta$ by maximizing the likelihood in (\ref{eq:4.3}), the Expectation-Maximization (EM) algorithm \citep{dempster1977maximum} is the standard optimization algorithm.}

Some other methods, including \texttt{WemIQ} \citep{zhang2014wemiq}, \texttt{Salmon} \citep{patro2017salmon}, \texttt{iReckon} \citep{mezlini2013ireckon}, and \texttt{MSIQ} \citep{li2016msiq}, introduce hidden variables to denote \JJL{the isoform origins of reads and use} these variables to simplify the form of the \JJL{likelihood function}. Suppose that the isoform origins \JJL{of reads $\bm R = \{r_1,\ldots,r_n\}$} are denoted as $\bm Z = (Z_1, Z_2, \dots, Z_n)^T$, where $Z_i = j$ if read $r_i$ comes from isoform $j$. Then the joint probability \JJL{density} of $\bm R$ and $\bm Z$ can be written as
\begin{align}
\begin{split}\label{eq: lik_read}
\JJL{p}(\bm R, \bm Z|\bm \Theta)&= \prod_{i = 1}^n \JJL{p}(r_i, Z_i|\JJL{\bm\Theta})\\
&= \prod_{i = 1}^n \prod_{j = 1}^J \left[\JJL{p}(r_i|\text{isoform}\ j)\;\theta_j\right]^{\JJL{\1}\{Z_i = j\}}\,. 
\end{split}
\end{align}
\JJL{Such model formulation is} especially useful when one would like to \JJL{estimate $\bm \Theta$} under the Bayesian framework \w{(Figure \ref{fig: txlevel}c)}, as \JJL{what has been done} in \texttt{MISO} \citep{katz2010analysis}, \texttt{Salmon} \citep{patro2017salmon}, and \texttt{MSIQ} \citep{li2016msiq}. Prior knowledge on $\bm \Theta$ can be incorporated \JJL{via modeling the prior distribution of $\bm \Theta$, and $\bm \Theta$ would be estimated as the maximum-a-posteriori (MAP) estimator.}
\w{As shown in Figure \ref{fig: txlevel}c, another advantage of the Bayesian framework is that the model can be easily extended to incorporate multiple RNA-seq samples and borrow isoform abundance information across samples \jl{\citep{li2016msiq}}.}

\add{A more recent isoform quantification method \JJL{\texttt{alpine}} \citep{love2016modeling} belongs to the third fragment-based category. \texttt{Alpine} is specifically designed to adjust for multiple sources of sequencing biases in isoform quantification. It considers all potential fragments with lengths within the middle of the fragment length distribution, at all possible positions within every isoform. For each fragment, \texttt{alpine} counts the number of reads compatible with it. Then \texttt{alpine} models the fragment counts using a Poisson generalized linear model, whose predictors are bias features including the length, the relative position, the read start sequence bias, the GC content and the presence of long GC stretches within every fragment. \texttt{Alpine} estimates the read start sequence biases using the variable length Markov model (VLMM) proposed by Roberts et al \citep{roberts2011improving} and implemented in Cufflinks \citep{trapnell2010transcript}. After estimating bias parameters, \texttt{alpine} outputs bias-corrected isoform abundance estimates. The Poisson parameter $\lambda_s$ for a potential fragment $s$ is assumed to be $\lambda_s = \sum_{j=1}^J a_{sj} \theta_j$, similar to what is assumed in region-based models. Hence, $\theta_j$'s are estimated based on the bias-corrected estimates $\hat\lambda_s$'s.}

The above \JJL{approaches, however,} would not lead to \JJL{accurate isoform} reconstruction results when directly \JJL{used} to discover new isoforms, because the number of candidate isoforms can be huge when the number of exons is large. A common practice is to add penalty terms before maximizing the objective function\JJL{, i.e., the likelihood or the posterior}. The regularization aims to enforce sparsity \JJL{on the} estimated $\hat{\bm \Theta}$, whose nonzero entries indicate the discovered isoforms. Two such reconstruction methods are \texttt{iReckon} \citep{mezlini2013ireckon} and \texttt{NSMAP} \citep{xia2011nsmap}.

\subsection{Regression-based \JJL{methods}}
The second type of statistical methods for isoform quantification is \JJL{regression-based}. These \JJL{methods formulate the isoform quantification problem as a linear or generalized linear model and treat the region-based read count (or proportion) as the response variable, candidate isoforms as predictor variables, and isoform abundances as coefficients (parameters) to be estimated}. Regression-based methods include \texttt{rQuant} \citep{bohnert2010rquant}, \texttt{SLIDE} \citep{li2011sparse}, \texttt{IsoLasso} \citep{li2011isolasso}, and \texttt{CIDANE} \citep{canzar2016cidane}. 

The basic model \JJL{is a linear model with region-based read count proportions as the responses}.
As for the design matrix, \texttt{IsoLASSO} considers a binary matrix to denote the compatibility between the isoforms and \JJL{genomic regions (i.e., a value of $1$ indicating that an isoform and a region are compatible, and $0$ otherwise)}, while the other three methods consider a conditional probability matrix\JJL{, for which the read proportions are} modeled as:
\begin{align}
\begin{split}
\frac{X_s}{n} &= \sum_{j=1}^J \JJL{\Prob}(\text{\JJL{a random read falls into} region}\ s| \text{isoform}\ j)\JJL{\;\Prob}(\text{isoform}\ j) + \epsilon_s\\
&= \sum_{j=1}^J F_{sj}\;\theta_j +\epsilon_s, \qquad s\in S\,,
\end{split}
\end{align}
where $\epsilon_s$ represents independent random noise with mean $0$. 
\w{As in the likelihood-based \jl{methods}, the probability $F_{sj}$ depends on the structure of region $s$ and the length of isoform $j$. Especially when region $s$ spans \jl{alternative} splicing junctions \jl{(e.g., region $s$ skips the middle exon but includes the two end exons)}, \jl{the} estimation \jl{accuracy} of $F_{sj}$ is \jl{critical} in the modeling.}
Then the estimation task reduces to a \JJL{penalized} least-squares problem
\begin{align}
\begin{split}
\hat{\bm \Theta} &= \argmin_{\bm \Theta \geq \bm 0}\sum_{s=1}^S\left(\frac{X_s}{n} - \sum_{j=1}^J F_{sj}\;\theta_j\right)^2 + \text{penalty},
\end{split}
\end{align}
where the penalty term is \JJL{only needed for isoform discovery and often excluded for isoform quantification}. For example, \texttt{IsoLASSO} sets the penalty term as $\lambda\sum_{j=1}^J\frac{n\theta_j}{L_j}$, where $L_j$ is the length of isoform $j$, while \texttt{SLIDE} uses $\lambda\sum_{j=1}^J\frac{\theta_j}{m_j}$, where $m_j$ is the number of exons in isoform $j$. \JJL{For both methods, $\lambda$ is a tuning parameter to control the level of regularization.} \texttt{IsoLASSO} selects $\lambda$ based on the resulting number of isoforms with non-zero \JJL{estimated} expression, while \texttt{SLIDE} uses a stability criterion \JJL{\cite{meinshausen2010stability}}.

\begin{remark} 
There are isoform discovery methods that reconstruct mRNA transcripts based on deterministic \JJL{graph} methods. Examples include \JJL{a} \textit{de novo} approach \texttt{Trinity} \citep{grabherr2011full}, and \JJL{reference genome-based} approaches \texttt{Scripture} \citep{guttman2010ab}, \texttt{Cufflinks} \citep{trapnell2009tophat}, and \texttt{Stringtie} \citep{pertea2015stringtie}\JJL{,
which all construct} splice graphs based on aligned reads and then use various criteria to parse the constructed graph into transcripts \JJL{in a deterministic way, without resorting to statistical models}.
\end{remark}

\begin{remark}
\w{Despite \jl{many methods developed} for isoform quantification, not all of them discuss the estimation uncertainty of isoform abundance levels. Even though the \jl{point estimates of} expression levels \jl{have led to new scientific discoveries in} many biological studies, it is \jl{important} to consider estimation uncertainty, \jl{especially when the differential expression analysis is of interest, or when some candidate isoforms are highly similar in structures (related to the collinearity issue in linear model estimation)}.
One way to evaluate the uncertainty in Bayesian methods is to construct posterior or credible intervals of the \jl{estimated abundance} levels \citep{li2016msiq,wang2010isoform}.
In regression-based methods, it is possible to calculate the standard errors of the abundance estimates (the coefficients in regression models). \jl{However, we have to note that assumptions, which are not always practical, are needed for uncertainty estimation. This explains why hypothesis tests about the same population abundance levels can give different $p$-values when they use different assumptions}.}
\end{remark}

\begin{remark}
There have been multiple efforts to quantify transcripts \JJL{for better accuracy} based on multiple \w{RNA}-seq samples (especially biological \JJL{replicates}), thanks to reduced sequencing costs and \JJL{the rapid accumulation} of publicly available RNA-seq samples. Model-based methods include \texttt{CLIIQ} \citep{lin2012cliiq}, \texttt{MITIE} \citep{behr2013mitie}, \texttt{FlipFlop} \citep{bernard2014efficient}, and \texttt{MSIQ} \citep{li2016msiq}. 
These methods generalize the models designed for \JJL{isoform quantification based on a single sample, and their results} show that aggregating \JJL{the} information from multiple samples can \JJL{achieve better accuracy} in isoform abundance estimation.
It has been noted in \JJL{\texttt{MSIQ} \cite{li2016msiq} that it is important to consider the possible heterogeneity in the quality of different samples} to obtain robust and accurate \JJL{estimation} results.
\end{remark}

\begin{remark}\label{remark:tx_compare}
Current statistical methods differ in their perspectives to formulate the isoform quantification problem, the trade-offs between the complexity and flexibility of models, and the methods to adjust for various sources of sequencing biases and errors. Because of the complexity of transcript-level analysis and the \JJL{noise and biases} in RNA-seq samples, it is impossible to identify a superior method for all real datasets. We suggest that users consider their preferences on the precision and recall rates in discovery problems, and to evaluate \JJL{the assumptions of different methods for RNA-seq read generation and bias correction}, before selecting the tool to apply on their data. For a computational comparison of some methods mentioned above, please refer to \cite{kanitz2015comparative} and \cite{steijger2013assessment}.
\end{remark}

\section{Exon-level analysis: exon inclusion rates in alternative splicing}\label{sec:exons}
Since transcript-level analysis of complex genes in eukaryotic organisms remains a great challenge \citep{steijger2013assessment}, there are approaches focusing on exon-level signals, seeking to study alternative splicing based on exons and exon-exon junctions instead of full-length transcripts. When transcriptomic studies focus on the exon-level, a primary step is usually to estimate the \textit{percentage spliced in} (PSI or $\Psi$, \citep{katz2010analysis}) of \JJL{an exon} of interest. Our discussion below applies to \JJL{an individual exon}. Considering two isoforms, one includes the exon and the other skips the exon, the goal of model-based methods is to estimate 
\ww{
\begin{align}
\begin{split}
\Psi &= \text{exon's inclusion rate}\\
&= \frac{\text{fraction of the inclusion isoform}}{\text{fraction of the inclusion isoform} + \text{fraction of the exclusion isoform}}\\
&= \frac{\JJL{\Prob}\text{(\JJL{a random} read is from the inclusion isoform)}}{
\JJL{\Prob}\text{(\JJL{a random} read is from the inclusion isoform)} +
	\JJL{\Prob}\text{(\JJL{a random} read is from the exclusion isoform)} 
}\,. 
\end{split}
\end{align}
}
A direct \JJL{estimator} pf PSI is
$$\hat\Psi = \frac{\frac{C_I}{L_I}}{\frac{C_I}{L_I} + \frac{C_E}{L_E}}\,,$$
where $C_I$ denotes the number of reads supporting \JJL{the inclusion isoform} (e.g., reads \JJL{spanning} the upstream splicing junction, the exon of interest, and the downstream splicing junction), \JJL{and} $C_E$ denotes the number of reads supporting \JJL{the exclusion isoform} (e.g., reads \JJL{spanning parts of the upstream and downstream exons but skipping the exon of interest}). $L_I$ and $L_S$ denote the lengths or \JJL{the} adjusted lengths (\JJL{after accounting for constraints on read and isoform lengths, i.e., isoform lengths $-$ read length}) of the inclusion and exclusion isoforms.


To evaluate the estimation uncertainty, methods including \texttt{MISO} \citep{katz2010analysis}, \texttt{SpliceTrap} \citep{wu2011splicetrap}, and \texttt{rMATS} \citep{shen2014rmats}  \JJL{use} different statistical models. Both \texttt{MISO} and \texttt{SpliceTrap} construct \JJL{models similar to the} model (\ref{eq: lik_read}) under the Bayesian framework, with $\Psi$ as the parameter of interest. Bayesian confidence intervals of \JJL{$\Psi$} can then be obtained based on its posterior distribution. \texttt{rMATS} accounts for the information from multiple replicates through the following hierarchical model
\begin{align}
\begin{split}
C_{Ik}|\Psi_k &\sim \text{Binomial}(n = C_{Ik} + C_{Ek}, p = f(\Psi)), \\
\text{logit}(\Psi_k) &\sim \text{Normal}(\mu = \text{logit}(\Psi), \sigma^2),
\end{split}
\end{align}
where $C_{Ik}$ ($C_{Ek}$) is the number of reads supporting inclusion (exclusion) in replicate $k\ (k = 1,2,\ldots,K$); $\Psi_k$ is the \JJL{PSI of the exon of interest} in replicate $k$; $\Psi$ and $\sigma^2$ are the mean and variance of PSI in the biological condition of interest; $f$ is \JJL{a} function to normalize $\Psi$ \JJL{based on the effective length of the exon}. Since both \JJL{\texttt{MISO} and \texttt{rMATS}} can estimate $\Psi$ and \JJL{the uncertainty of $\hat\Psi$}, it follows that they can detect differential exon usage between two biological conditions through statistical testing.

\begin{remark}
 The above discussion mainly focuses on the scenario\JJL{, where only two alternative isoforms are involved,} and does not extend easily to more complex alternative splicing patterns with more than two alternative splice forms. A proposed remedy is \texttt{DiffSplice} \citep{hu2012diffsplice}, which identifies alternative splicing modules (ASMs) from the splice graph to study splicing patterns that may involve multiple exons. However, one limitation of \texttt{DiffSplice} is that it does not address the estimation uncertainty of the expression levels of ASMs.
 \ww{\texttt{DEXSeq} \citep{anders2012detecting} is another method that studies differential exon usage, but it focuses more on exon-level expression and less on splice junctions. }
\end{remark}

\begin{remark} 
There is a trade-off in alternative splicing studies concerning whether to use transcript-level or exon-level \JJL{information}. Full-length transcripts provide global information on splicing patterns which directly lead to knowledge on protein isoforms, but accurate quantification of transcripts suffer from the limited information in short RNA-seq reads. 
On the other hand, exon-level analysis results in the more accurate quantification of individual splicing events, but limit the scope of studies to local genomic regions. As mentioned in Section \ref{sec:txs}, the accumulation of multiple RNA-seq samples and the \JJL{increasingly large databases} of annotated transcripts \citep{harrow2012gencode} might provide a solution to this dilemma: \JJL{combining information from multiple samples with} prior knowledge on \JJL{transcripts} may assist the reconstruction and quantification of \JJL{full-length} isoforms from \JJL{short RNA-seq reads}.
\end{remark}

\section{Outlook}\label{sec:discussion}
RNA-seq has become the standard \JJL{experimental} method for transcriptome \JJL{profiling}, and its application to numerous biological studies have led to new scientific discoveries in various \JJL{biomedical} fields. We have summarized the key statistical considerations and methods involved in gene-level, transcript-level, and exon-level RNA-seq analyses.
Despite the fact that continuous efforts on the development of new tools improve \JJL{the} accuracy of analyses on all levels, challenges posted by relatively short RNA-seq reads remain \JJL{in studying} full-length transcripts, making it difficult to fully understand the dynamics of \JJL{mRNA} isoforms and their protein products. 
In complex transcriptomes, probabilistic models have limited power in distinguishing different but highly similar transcripts.
It has been noted that identification of all constituent exons of a gene is not always successful, and in cases where these exons are correctly reported, it is challenging to assemble them into complete transcripts with high accuracy \citep{steijger2013assessment}.
Given the current read lengths in NGS, we emphasize the importance of jointly using multiple samples (i.e., technical or biological replicates) to aggregating information on \JJL{alternative splicing and sequencing noise}. \JJL{Na\"ive} pooling or averaging methods are shown to be inadequate in the multiple-sample analysis \citep{li2016msiq}, and statistical discussion on this topic is still insufficient.
On the other hand, new sequencing technologies such as PacBio \JJL{\cite{rhoads2015pacbio}} and Nanopore \w{\cite{branton2008potential,byrne2017nanopore}} sequencing technologies can produce longer reads with average \JJL{lengths} of $2-3$ kb \citep{au2013characterization}. A \JJL{primary barrier} of the current long-read sequencing \JJL{technologies is their relatively high error rates and sequencing costs} \citep{bleidorn2016third}. \w{One current approach to take advantage of these new technologies is to combine the information in next-generation short reads and third-generation long reads in isoform analysis \citep{au2013characterization}.
}

\w{To demonstrate the efficiency of statistical methods developed for RNA-seq data, method developers must show the reproducibility and interpretability of these methods. As we have discussed in \textit{Remark} \ref{remark:de_compare} and \ref{remark:tx_compare}, there is hardly a method that is superior in every application. However, a useful method should at least demonstrate its advantage under specific assumptions or on a particular type of datasets. Meanwhile, no matter how complicated a statistical model is, its general framework and logical reasoning should be interpretable to the users (e.g., biologists).
Also, comparison of different methods over benchmark data can be beneficial for the development of new methods. Experimentally validated benchmark data for RNA-seq experiments are still limited on the genome-wide scale.
}

Aside from the \JJL{analysis tasks} introduced and discussed in this \JJL{review article}, RNA-seq is also widely applied to other areas like RNA-editing analysis \citep{ramaswami2012accurate,bahn2012accurate}, non-coding RNA discovery and characterization \citep{iyer2015landscape,hezroni2015principles}, expression quantitative trait loci (eQTLs) mapping \citep{pickrell2010understanding}, 
\w{and prediction of disease progression \citep{zak2016blood},}
with interesting statistical questions involved.
\w{Transcriptomic data can also be integrated with genomic and epigenomic data to \jl{advance our} understanding of gene regulation and other biological processes \citep{hawkins2010next}.}
In recent years, the emerging single-cell RNA sequencing (scRNA-seq) technologies enable the investigation of transcriptomic landscapes at \JJL{the} single-cell resolution, \JJL{bringing} RNA-seq analyses to a new stage \citep{kolodziejczyk2015technology}. 
\JJL{In contrast to scRNA-seq data, the RNA-seq data we have reviewed in this article are now referred to as bulk RNA-seq data, where the data are generated from RNA molecules in multiple cells in a batch.} The analysis of scRNA-seq data is complicated by excess zero counts, the so-called dropouts due to the low amounts of mRNA sequenced within individual cells. Therefore, current usage of scRNA-seq data \JJL{focuses} on gene-level analysis\JJL{,} and frequently discussed statistical topics include clustering \citep{xu2015identification}, dimension reduction \citep{pierson2015zifa}, and imputation \citep{li2018accurate}. Since the signal-to-noise ratio in scRNA-seq data is much lower than that in bulk RNA-seq, many models developed for bulk RNA-seq \JJL{data} cannot be directly applied to scRNA-seq \JJL{data}, calling for the development of new computational and statistical tools. 
\w{With the ongoing efforts to build the Human Cell Atlas \cite{regev2017human}, new scRNA-seq and other single-cell level data \jl{(e.g., imaging data)} will help people more thoroughly understand human cell types and their \jl{molecular mechanisms}. People can also refer to The Human Cell Atlas White Paper for the detailed discussion of statistical challenges in analyzing these data \citep{HCA}.}

\clearpage
\bibliographystyle{unsrt}
\bibliography{rnaseq}

\end{document}